# Generating relevant scenarios for intelligent transportation service

Ismet Addoui [1,a], Tarek Chouaki [1,2,a] and Ambrogio Delli Colli [1,3,a]

[1] *IRT SystemX, 8 Avenue de la Vauve, 91127 Palaiseau Cedex, France*
[2] *Sorbonne University, 4 Place Jussieu 75005 Paris, France*
[3] *Politecnico di torino, Corso Duca degli Abruzzi, 24 10129 Torino, Italy*

**Abstract.** This paper addresses risk assessment issues while conceiving complex systems. Indeed, project stakeholders have to share the same problems understanding allowing to undertake rational and optimal decisions. We propose an approach based on Natural Language Processing (NLP) techniques to improve systems quality requirements such as consistency and completeness. We assess the relevancy of our approaches through experimentations and highlighted feedbacks from project stakeholders and players.

## 1 Introduction

Requirements Engineering aims to define documents and maintain requirements [1]. The number and nature of activities characterizing any process for managing requirements is correlated with the target domain (software, real-time embedded systems, automotive or railway), as well as the importance of the project, its innovative nature and other considerations such as human factors, standards or technologies. Any requirements engineering process includes at least activities of elicitation, specification, validation and change management. These activities are supported by different techniques such as interviews, knowledge engineering, modelling and analysis or simulation.

Having a clear understanding of the customer's needs is not an easy task. The goal of the requirements engineer is to help customers to clarify and to frame the requirements of the system under construction. Building a complete, consistent and feasible requirements base is a challenging issue and a timely research topic. Moreover, the quality of the requirements is of a high importance to avoid snowball effects leading to misinterpretation and ineffective realizations.

Without a shared vision, chances to converge quickly to a requirement base that is complete, consistent and feasible are negligible or would require many iterations leading to higher costs, delays and competitiveness. This comes besides with no guarantee on the quality of the requirements. To improve the quality of the requirements, one can consider scenarios with important advantages already addressed in the literature [2, 3]. For instance, scenarios tend to adopt the user's viewpoint, which is a key feature to validate the adequacy between the requirements and the user's needs. Considering the complexity of large systems, the number of scenarios needed to effectively help practitioners to design a system that fits the needs of the customers can be important. The selection of relevant scenarios among possible ones to cover the user's need is not obvious. Today, this task is generally performed in ad-hoc manner by experts that consider the system only through their prism, generally limited to their concerns. We think there is a need to improve this situation and to generate more relevant scenarios considering the full coverage of users' needs and required analysis to be performed to validate their adequacy, their completeness or their feasibility.

[a] Corresponding author : firstname.lastname@irt-sytemx.fr

## 2 Motivations and our contribution

Several studies highlighted recurrent problems that occur when large scale projects fail. For instance, the CHAOS Report [4] lists the main reasons IT projects fail. We can cite among them:

- Bad vision / understanding of the users' needs
- Lack of implication of the various stakeholders early in the development life cycle
- Difficulties for the stakeholders to collaborate (silos between experts)
- Lack of relevant metrics to take rational decisions

In practice, most of problems issued from bad decision making are discovered during verification and validation activities. When the first versions of the system have been implemented and deployed. Those late discoveries require most of time re-engineering the system at several levels: operational, functional or physical; or reconsidering the strategy of the enterprise in terms of partnerships or skills to be developed.

From a technical perspective, the design of a complex system driven by scenarios can be helpful if it allows the consideration of any relevant combination of features (events, behaviours, conditions) to really get a deeper understanding of how the system should behave and what it should be made of. The question we shall consider are as follows:

- How many scenarios do we need to consider to embrace all user's needs?
- What are the combinations of events, behaviours or conditions do we need to analyse in priority according to initial hypotheses?

The previous questions are important and necessitate detailed responses. Nevertheless, there are no explicit answers and we propose some experiments with prominent preliminary results providing some hints.

## 3 Application

We aim to design and implement a mobility service where systems must cooperate to fulfil the expected operation capabilities. We consider five capabilities described in the following:

- Enhance personal mobility: this can be measured through user interviews about comfort or quality of service. Related metrics are the average time of waiting to access the transportation service or the mean time delay to reach the desired destination.
- Ensure the System of System (SoS) integrity: it is about the physical integrity of any human being involved in operational scenarios (user, pedestrian, cyclist, etc.). This latter could be corrupted in many ways: physical attacks on roadside units or vehicles, or cyber-attacks. Meeting this capability can be measured by the reduction of any kind of accidents. This implies also a good crisis management in case of accident, in order to not deteriorate the situation and not create congestions.
- Improve environment impact: this is related to the transportation efficiency and effectiveness of the cooperation between vehicles and roadside units. Meeting this capability can be measured by the reduction of fuel consumption or lower emission of $CO_2$ fine particles.
- Ensure service continuity: this allows to avoid malicious acts that could lead to corruption of human beings, vehicles or roadside units' integrities. It also tries to anticipate damages. Its satisfaction can be measured by the reduction of malicious acts or their associated costs.
- Reduce of congestion. It can be measured by the reduction of occupation rates of roads, or by the reduction of annual mean time spent on the roads by users.

Our analysis is driven by usage scenarios leading to ask the following questions:
- How systems implied in the scenarios shall behave and cooperate to fulfil each of those operational capabilities?
- What are the data they shall exchange? Through which media and what quality of service?

Depending on a described situation, the set of behaviours and interactions that shall be considered can be very large. Hence, this cannot be considered and managed easily at design time. Nevertheless, in this paper, we propose to generate a set of representative scenarios to cover a maximum number of cases, leading to a consistent and complete specification. The quality and the accuracy of simulations' results of the generated scenarios have an important impact on the ability to proceed to relevant architecture choices.

## 4 Generation of scenarios

Figure 1 depicts and illustrates our approach to generate scenarios. Note that, the definition of scenarios considered is "sequence of events that occurs during a particular execution of a group of objects implied in the system of system". The set of scenarios to drive the analyses at system level, is performed by experts.

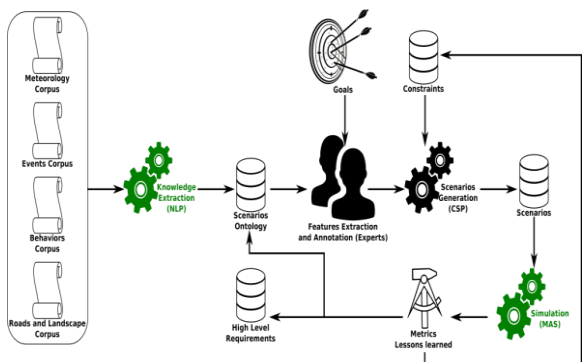

*Figure 1: Overview of the scenarios generation*

This is a manual activity relying on the knowledge of these experts and their ability to imagine key scenarios to validate users' needs. Pragmatically, the required number as well as the relevancy of scenarios defined by our experts cannot be measured accurately. Knowing that this set of scenarios is produced by humans, we can assume that there are omissions due to the large scope of features. For instance, experts have defined over 50 scenarios ranging from the highest probability of occurrence to the most critical. Are those scenarios enough to cover all the needs? Considering the complexity of the SoS, the answer is obviously no. Then there is a need to assist expert to elicit more relevant scenarios.

### 4.1 Knowledge Extraction

To improve the quality and the accuracy of scenarios driving key decisions, we have used an approach using knowledge extraction from the dedicated literature in order to capitalize it into an ontology. This ontology classifies all key features characterizing any scenario: users, vehicles, roadside units, road types or traffic conditions. For instance, our ontology describes all kind of:

- Climatic conditions: brightness, rain, snow, wind, fog
- The various roads the vehicles can take; sloppy, curving, muddy or icy
- The taxonomy of all behaviours that can be expected from drivers or pedestrians: careful, distracted or dangerous

This knowledge has been extracted from a set of corpuses using Natural Language Processing with the NLTK framework [5]. The obtained results using NLP are completed with expert recommendations.

As illustrated in Figure 2. We extract key terms from the documents of the mobility corpus after tokenizing, cleaning and lemmatizing content. We rely on the computation of semantic similarities using WordNet [6] to reduce the resulting set of words. Key terms extraction relies on the frequency distribution of remaining words. Finally, we use tags from POS (Part of Speech) Tagging to determine the role of corresponding words in order to generate predicates. These predicates are validated / completed by mobility experts.

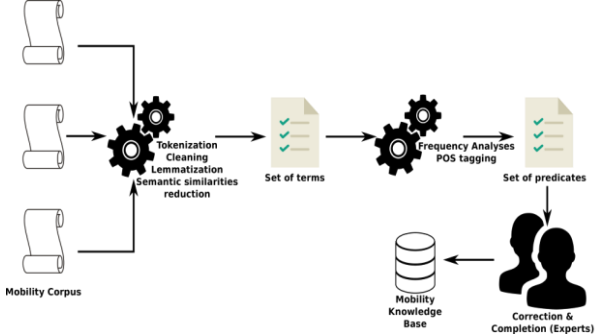

*Figure 2: Ontology extraction using NLP*

### 4.2 Generation of scenarios

The extracted ontology is exhaustive and can be used for a huge number of scenarios. Since our experimentation concerns a specific zone in Paris area, we extracted the relevant subset for achieving deeper analyses considering experts' recommendations. For this purpose, we relied on data we collected according to our experimentation: climate, hazards, topology of road, localization of roadside units, etc. We selected 45 features regarding the operational capabilities we wanted to reach. Then, each feature is annotated with a level of probability of occurrence ranging from A (the most probable) to E (the less probable). The features are also annotated with a level of criticality ranging from A (the most critical) to C (the less critical). Table 1 shows the list of selected features along with their respective levels of probability of occurrence and criticality regarding the scope of our experimentation.

**Table 1.** Scenario Features

| Features | Criticality | Probability |
|---|---|---|
| Curved Road | B | A |
| Speed Bump | B | C |
| Teleoperation Hacking | A | D |
| CEM | A | E |
| Cut Out | B | B |
| Equipements Supervision Hacking | A | C |
| Impaired Walking Pedestrian | A | C |
| Overtaking | A | A |
| Against Rally Driving | A | D |
| Safety Distance Violation | C | A |

In order to generate consistent scenarios, we propose to define some features' constraints. The two following formulas illustrate example of meet constraints.

$$\neg(cross\_road \wedge round\_about) \quad (1)$$
$$Fog \rightarrow Low\_grip \quad (2)$$

Formula (1) is used to avoid scenarios with both crossroads and roundabouts. Aware that (in our study) it is impossible to have simultaneously a crossroad and a roundabout.

Formula (2) is used to force the presence of low grip in case of fog.

Adding those constraints allow us to avoid having inconsistent scenarios.

It is important to mention that with 45 selected features and 22 specified constraints; the total number of possible scenarios remains very high (one can consider an exponential number of combinatorial cases scenarios).

Since we cannot afford to analyse all these scenarios, we have investigated a new approach to prioritize these scenarios using two metrics such as criticality and probability of occurrence. For this purpose, and for each generated scenario, we also propose two scores as indicated by equations (3) and (4):

$$P_g = \prod_{i=1}^{n} P_i \quad (3)$$
$$C_g = \prod_{i=1}^{n} 2^{C_i} \quad (4)$$

Equations (3) and (4) provide the global probability (resp. criticality) of a scenario for each of them using the probability of occurrence (resp. criticality) of each feature.

Equation (4) uses an exponential law as it fosters highest levels of criticality. Once a global score for each scenario has been computed in terms of probability of occurrence and criticality, we represent these scenarios in Figure 3.

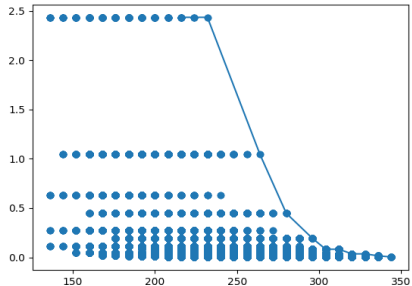

*Figure 3: Pareto front of the generated scenarios criticity over probability of occurence*

This figure represents the criticality's behaviour according to probability of occurrence variations.

Before going through the details of Figure 3, we propose the following definition [7]
The definition by [7] set of non-dominated solutions, being chosen as optimal, if no objective can be improved without sacrificing at least one other objective. On the other hand, a solution x∗ is referred to as dominated by another solution x if, and only if, x is equally good or better than x∗ with respect to all objectives. The Pareto front points (see figure 3) are the non-dominated solutions.

Figure 3 depicts the Pareto front allowing us to extract the most representative scenarios represented by non-dominated points according to the two identified criteria (criticality and probability of occurrence). This Pareto front is an efficient approach to considerably reduce the total number of scenarios and exhibits only the non-dominated points or scenarios that we consider the most important to be investigated. From this 1134 scenarios have been selected from the Pareto front for analysis and study. Items below represent features of a single scenario to be prioritized for analysis:

- Impaired Walking Pedestrians
- Right Overtaking

- Communication Loss
- Overtaking
- Safety Distance Violation
- Direction Change Not Communicated
- Wind
- Out of Zebra Crossing
- Over speed Driving
- Glare
- Equipment Supervision Hacking
- Cut Out
- Impaired Driver
- Late Obstacle Detection
- VRU Obstacle
- 2 Wheels
- Cut In
- Emergency Vehicles
- Low Grip
- Curved Road

Note that items of this example precise only the features that shall appear into the same scenario. They do not reveal information about the causality links between events or behaviours. To cope with this issue, we appealed experts to recommend and select the most relevant scenarios depicted by Pareto front approach.

For instance, we provide in the following an example illustrating a right interpretation of features into a realistic scenario:

*"The autonomous vehicle drives on a sloppy slippery road with low visibility because of the fog. Electromagnetic disturbances cause a loss of communication with the control / command center. The road taken is often crossed by wild animals."*

Such situation can cause the immediate stop of the transportation service otherwise, it can put the life of users in danger. In this case, we need a set of functions ensuring the good coordination between vehicles and roadside units. This situation requires also a function for the vehicle to park in a safe area and notify the control-command operator as soon as the communication is fully restored. This analysis indeed impacts behaviours and interactions required to ensure safer transportation services.

## 5 Conclusion

In this paper, we presented some achievements of a novel approach making use of techniques from AI (NLP for instance) to elicit requirements for a mobility service. Compared to traditional approaches, we took the advantage of AI techniques to help practitioner to focus on the most important features when designing a complex system. This approach allows us to select the nature of the scenarios needed. The generation of scenario could focus on the criticity, the probability of occurrence or both. The feedback we had from the experts is considered to improve the quality of the obtained results. Besides, most of techniques presented in this paper have been automated (knowledge extraction, scenario generation, exploitation of results) which is a key point for dynamically managing changes. Obviously, our approach suffers from several limitations in terms of techniques or application that shall be addressed by further works. For instance, we did not take account the strategic aspects of the enterprise into the elicitation of requirements and we do not generate the story of the scenario. Finally, the temporality of the appearance of event (features) is not specified in our method.